# A Broader Perspective about Organization and Coherence in Biological Systems[1]

Martin Robert

Institute for Advanced Biosciences, Keio University, Tsuruoka, Yamagata, Japan
mrobert@ttck.keio.ac.jp

**Abstract.** The implications of large-scale coherence in biological systems and possible links to quantum theory are only beginning to be explored. Whether quantum-like coherent phenomena are relevant, or even possible at all, at the high temperatures of biological systems remains unsettled. Here, we present a broader perspective on biological organization and how quantum-like dynamics and coherence might shape the very fabric from which complex biological systems are organized. Regardless of its exact nature, a unique form of coherence seems apparent at multiple scales in biology and its better characterization may have broad consequences for the understanding of living organisms as complex systems.

## 1 Background

The spatio-temporal organization, coherence and complexity found in biological systems appear overwhelming and their underlying principles are only beginning to be deciphered. It is widely accepted that these principles must be based on more fundamental ones originating at the sub-molecular or molecular level. Since life depends on complex networks of chemical reactions, there are likely deep connections to the underlying quantum theory, which defines chemistry so well. However, quantum phenomena are often associated with paradoxical states characterizing the microscopic world of subatomic particles or other fundamental levels of matter that do not easily fit our perception of the reality of larger systems, including biological ones. In addition, while quantum theory is considered to be one of the most successful scientific theories, its implications beyond the microscopic level are only beginning to gain attention. Whether quantum-like phenomena are relevant or even possible at all, at the high temperatures of biological systems remains somewhat controversial. Here, based on existing findings and proposals we explore its possible relevance and utility in providing insight into the remarkable dynamics and coherence observable in living systems. We highlight some specific examples of organization and large-scale coherence in biological systems that might be connected or display properties better described by non-classical or quantum-like features including quantum coherence.

## 2 Proposal/results.

### 2.1 Mechanisms to maintain coherence

The standard reasoning suggests that quantum coherence can't be maintained at high temperature in biological systems for any significant period of time due to strong coupling to the environment. Prior studies have shown that under conditions relevant in biology, systems will decohere very rapidly [1]. However, some recent studies provide evidence that quantum coherence can be maintained, in some systems of biological relevance, for much longer time-scales than expected [2]. Perhaps the best established example supporting the presence of quantum coherence in biological systems include electron transfer machinery in photosynthesis [3, 4]. Other suggestive examples where direct evidence may still be lacking include the avian magnetoreception and orientation system, enzyme catalysis and reaction dynamics [5], and some effects in ligand-receptor interactions in cellular signaling processes [6]. In addition, a quantum mechanical model has been proposed as a possible mechanism to explain

---

some forms of adaptive mutation [7]. However, most of these examples of possible quantum coherence in biological systems appear to be extensions of widely accepted quantum chemistry principles, albeit occurring within the warm environment of the cell. As such, these processes may not be so surprising. Here, we aim to reflect mainly on whether quantum-like phenomena might play a role or have explanatory power at higher levels of organization including the cellular, organ, and even whole organism level.

**1) Open confinement.** Arguments for ruling-out any form of quantum-like coherence at high temperature often seem to ignore some important and unique properties of biological systems. These include prominently, the confined, protected, and controlled cellular environment, a product of the insulating nature of the cell membrane. This complex structure provides spatial confinement of internal elements and activities while at the same time allowing the necessary channels to operate two-way exchanges with the environment. The cell thus constitutes a unique spatially confined system with an open architecture that facilitates controlled exchanges with its environment. This "open confinement" appears critical in maintaining cellular coherence. It may seem paradoxical that coherence could be maintained through active exchanges with the environment, the very phenomenon that is usually considered the main factor in decoherence. However, the type of exchanges occurring across the living cell membrane, for example, are highly specific and are controlled by selective molecular filters that allow only certain exchanges, in contrast to other systems. In addition, specific forms of interactions and coupling between a system and its environment have been shown to be coherence-promoting [1].

**2) Metabolic activity and negentropy.** In addition, biological systems are known to operate in a thermodynamic state that is far from equilibrium by making use of energy consuming and energy producing metabolic activities. This allows for the constant renewal of cellular components and can be conceptualized as an important error correction mechanism and as countering decoherence. Without metabolism, cellular activities would soon come to a halt due, among other things, to the inability to replace defective cellular components and maintain a constant supply of energy. Schrödinger introduced the term negative entropy [8] or negentropy, linking it to metabolism, to describe this generation and maintenance of order in biological systems that are otherwise immersed in decoherence-inducing environments. This concept of negentropy therefore appears central in describing the constant renewal of damaged cellular constituents linked to normal cellular activity and how it acts to resist the natural tendency toward decoherence and to counter the unavoidable entropy generated by living systems.

**3) Multi-level communication and downward causation.** Finally, the complex and multi-scale forms of information exchange and signaling within biological systems further contribute to the maintenance of long-term coherence, regardless of the exact description that this coherence may fit. As argued by Noble [9, 10], organization in biological systems cannot be simply described from bottom-up principles but is strongly linked to downward causation. In this form of interaction, elements that are the product of the whole system or that are generated from the higher levels of organization feedback and regulate lower level elements in a self-referential manner. Accordingly, there seems to be no privileged level of causation in biological systems and these levels interact at all scales. The self-generated feedback loop may be considered as strongly coherence-inducing mechanisms since the information is self-generated, self-reinforcing and to some extent independent on external factors. It may thus conceptually also represent another form of system insulation against decoherence. Together these unique properties, and likely others not described here, provide a unique and dynamic infrastructure that allows the maintenance of a high level of organization and a possibly unique form of coherence in biological systems.

**2.2    On complex systems and organization**

Defining exactly what a complex system is remains a somewhat fuzzy issue. However, key features that have been associated with complexity include a system's openness with fuzzy boundaries, the presence of a nested architecture, memory or hysteresis, non-linearity and feedback loops, and emergence. Here, we mainly refer to the organized complexity connected with emergence. A basic consensus appears to be that complexity originates from systems with a large number of interacting elements. Moreover, as described above, it is likely that complexity is to some extent the result of downward causation, where the higher level emerging functionality and the associated information - which can't be intuitively deduced from the lower level component and its interaction - can feedback and regulate the system's elements at the lower level. This multi-level and intertwined bottom-up and top-down exchange of information may be characteristics of complex systems and particularly biological ones.

## 2.3 Coherence, unity, and the dynamics of complex systems

A high level of coherence is apparent in living systems and cellular processes taking the form of naturally occurring oscillatory phenomena. These oscillations appear linked to self-organized collective behavior and to the temporally organized expression of genes, proteins, and metabolites in single cells as well as in coherent cell populations. Striking examples of such high coherence include the oscillatory dynamics of most intracellular components observable in synchronized continuous cultures of yeast [11, 12]. Such continuous cultures of yeast, and also *E. coli*, have been shown to self-organize, resulting in synchronized respiratory oscillations associated with high temporal coherence at the level of gene, protein, and metabolite expression. While the exact mechanisms of synchronization have not been clarified, the cellular redox state as well as $H_2S$ and acetaldehyde levels have been found to be important [12–14]. In *E. coli*, valine and other metabolite exchanges might be responsible for similar synchronization events [15]. Such results suggest that most, if not all, cellular components oscillate, claims supported by other studies [16] and in other systems [17]. The phase and frequency of oscillations depends on conditions and can be manipulated. Analysis of their dynamics upon perturbations and over extended time periods show multi time-scale fractal-like structure and chaotic attractors [18]. Moreover, it is interesting to note that other dynamical cellular states have been characterized as chaotic attractors and proposed to be connected with cellular pluripotency and differentiation [19]. In addition, support for the relevance of quantum-like coherence in biological systems operating at the edge of quantum chaos [20] suggests there might be important analogies in such chaotic cellular systems that are worth exploring further. Experimental systems such as synchronized populations of yeast, or similar ones, might represent tools to explore these chaotic dynamical states. Extending similar arguments suggests that such phenomena may also be present at higher organizational levels, even the whole organism level, albeit at different spatio-temporal scales.

Another example to illustrate possible non-classical behavior may be a phenomenon reminiscent of quantum-like interference recently observed in bacterial gene expression [21]. These findings show how quantum-like statistical models may be useful, without being constrained by the absolute definitions derived from quantum physics. Basically, the authors argued that the dynamics of biological systems might require statistical descriptions where quantum-like probability amplitudes are more appropriate than classical probabilities. They also mentioned that this process might arise from the complexity of information processing in adaptive biological systems. Although the authors did not suggest that these quantum statistical models imply any underlying quantum processes, one could argue that it might be a likely conclusion.

Information is deeply connected with entropy, complexity, emergence and self-organization at multiple scales [22]. A recent report showed that some quantum computation models can result in descriptions of systems that have lower entropy than classical models [23]. Since biological systems display apparently complex but highly organized dynamics, such quantum-like model properties could play a role in minimizing entropy and maximizing information content. These authors also suggested that correlations between elements in complex systems may be connected to quantum effects and quantum information [24]. While their work was not presented or derived for this purpose, it may not be farfetched to try and extend this idea and to apply it to the highly correlated and coherent examples found in living systems, such as those described above. These principles might be connected with the encoding and compression of biological information and with the phenomenon of emergence. Is it possible that by extending existing concepts of chaos theory and complex systems, we might postulate that the coherent oscillatory phenomena that are omnipresent in biological systems constitute an informational infrastructure making quantum-like information processing and systems organization possible within cells, organs, and even whole organisms? The need to refer to non-classical behavior would arise when the complexity of a system is such that its behavior and interactions cannot be described by classical measures of coherence.

## 2.4 Limitations and need for further developments

It has often been argued that classical principles are sufficient to describe the form of coherence and non-linear dynamics that describe biological systems and many such classical approaches have been used successfully. This remains an important argument against the necessity to use quantum theory to describe biological systems, in addition to the fact that our usual experience of these systems does not seem to display the entanglement properties that typically characterize quantum systems. In addition, the non-linear causal structure in complex biological systems that may facilitate the maintenance of coherence may however be incompatible with quantum superposition, for example [25]. However, in spite of such arguments it remains important to ask whether the enormous phase space and computation-like tasks explored or performed by biological systems might require information processing that go beyond classical systems. Quantum search algorithms can allow to sense and explore multiple states simultaneously, a property that can provide efficiency that is beyond that made possible by classical systems [1, 3]. As we have described, the use of quantum mechanical principles at higher levels of biological organization may require further conceptual refinements that will account for both its relevance and limitations at this higher level of complexity.

The concept of unity and non-locality emerging in quantum states of matter like Bose-Einstein condensates and superconductors/superfluids, might be an important link or analogy to represent coherent organization in living systems, rather than more rigid definitions of quantum processes, based primarily on entanglement and superposition. Order in biological systems, such as within the cell, might produce functional analogs of such supercooled states of matter where a large number of particles are correlated and result in collective effects and the emergence of quantum phenomena at the macroscopic level. Such ideas about very low effective temperatures that may exist within the confined cellular environment have previously been explored for their relevance in biological systems [8, 26, 27]. In addition, a recent study revisited the phenomena of small numbers in biology, based on Schrödinger's work, and from this derived enticing ideas and inferences about the possible existence of quantum entanglement in cells [28]. The possible coherence maintaining mechanisms discussed here, and examples of coherent systems appear to complement well these ideas. In any case, some properties of biological systems might not be readily described or explained by classical theory, suggesting there may be some need to extend our view of quantum-like processes.

To this day, most evidence of quantum phenomena in biological systems is to be found at the microscopic, physical chemistry level [2, 6]. However as suggested here, it may also be at the higher level of cellular and organismal levels that quantum-like phenomena might play some unsuspected and significant role.

### 2.5 Testing a possible role of quantum-like processes in higher-level biological systems.

Overall, it seems there may be already some support both in existing experimental and theoretical studies that quantum-like phenomena have an unsuspected relevance in biological systems. Were clearer demonstrations of fundamental quantum behavior and even entanglement or superposition in biological systems to surface in the coming years, this would represent an exciting development. It may therefore be worth investigating further whether quantum-like coherent interactions may not only be relevant but might form an important element of the very fabric around which biological systems are organized. Because of the deep connections between non-linear dynamics and quantum properties in physical systems it should be possible to design experiments on dynamical cellular systems and determine, through their dynamical properties, whether quantum processes are at work at all. This could include obtaining further details and more definite measurements about the models of coherence in photosynthesis and enzyme kinetics, as described above. Alternately, these demonstrations could also take the form of accurate measurements of the dynamics of coherent cellular systems similar to those described here, or even the analysis of the thermodynamic exchanges of cellular/organism with their environment. Further research in this direction is therefore warranted.

### 3 Conclusion and significance

The issue of coherence in biological systems and its mechanisms remains unresolved and the debate about the relevance of quantum mechanics in biology continues. However, there is accumulating evidence that at some level, biological organization and the non-linear dynamics of cellular and organism may display properties highly reminiscent of quantum-like coherence that is not readily explained by classical means. It is also possible that the type of coherence displayed by biological systems fits neither the classical nor the quantum-like definition of coherence and might represent yet another form of coherence unique to living organisms. In any case, the significance of the presented view is not in trying to redefine what quantum-like coherent states may be in biological systems, but rather about the consequences of that coherence, whatever its nature. Such a broader perspective suggests that the key to the understanding of cellular and biological systems as inseparable wholes resides intrinsically in their non-linear and non-equilibrium dynamics as well as their highly coherent interactions. Ignoring these principles that are linked to self-organization and emergence is bound to result in failures in our attempts to understand and model living systems. Finally, this perspective on biological systems suggests a need for placing more emphasis, when observing, experimenting, and studying living and other complex systems, on the maintenance and proper analysis of the interactions between the system constituents at any level of complexity. Fortunately, such an objective seems to be at the core of a growing multidisciplinary field emphasizing systems approaches in biology and whose findings will likely have important consequences for biological sciences.

**Note:** This essay is meant to be conceptual and to stimulate interdisciplinary discussion and the author acknowledges that some of the ideas expressed remain hypothetical, especially with respect to the proposal of the relevance of quantum coherence at higher levels of biological organization. The main objective is to integrate existing concepts/knowledge and possibly apparently disparate points of view and reflect on whether they may be more connected than expected and how they might be useful for understanding complexity and coherence in biological systems.

**Acknowledgments.** The author is grateful to Johnjoe McFadden, Jaime Gómez Ramírez, and Takuya Morozumi for stimulating discussions and feedback.